\documentclass{article}

\usepackage{arxiv}

\usepackage[utf8]{inputenc} % allow utf-8 input
\usepackage[T1]{fontenc}    % use 8-bit T1 fonts
\usepackage{hyperref}       % hyperlinks
\usepackage{url}            % simple URL typesetting
\usepackage{booktabs}       % professional-quality tables
\usepackage{amsfonts}       % blackboard math symbols
\usepackage{nicefrac}       % compact symbols for 1/2, etc.
\usepackage{microtype}      % microtypography
\usepackage{graphicx}
\usepackage{natbib}
\usepackage{doi}

\title{Prior-mean-assisted Bayesian optimization application on FRIB Front-End tunning}

\author{ Kilean Hwang\thanks{\texttt{hwang@frib.msu.edu}}, Tomofumi Maruta, Alexander Plastun, Kei Fukushima \\ \textbf{Tong Zhang, Qiang Zhao, Peter Ostroumov, Yue Hao} \\
	Facility for Rare Isotope Beams\\
	East Lansing, MI, USA
}

\renewcommand{\headeright}{}
\renewcommand{\undertitle}{}

\hypersetup{
pdfsubject={acc.Phys},
pdfauthor={Kilean H. et. al.},
pdfkeywords={Prior mean, Bayesian optimization, Accelerator, Physics},
}

\begin{document}
\maketitle

\begin{abstract}
 Bayesian optimization~(BO) is often used for accelerator tuning due to its high sample efficiency. However, the computational scalability of training over large data-set can be problematic and the adoption of historical data in a computationally efficient way is not trivial. Here, we exploit a neural network model trained over historical data as a prior mean of BO for FRIB Front-End tuning.
\end{abstract}

% keywords can be removed

\section{Introduction}
 The Facility for Rare Isotope Beams (FRIB) at Michigan State University (MSU) is designed for various kinds of rare isotope production. This involves the frequent switch of the ion source species. Therefore, fast tuning of the accelerator Front-End (FE) to maintain optimal beam optics is one of the key performance requirements. Breaking-through the tuning performance over the traditional black-box optimization algorithm may be possible if historical or simulated data can be incorporated into the optimization algorithm in a computationally feasible way. However, we experienced significant machine status changes (a.k.a. `distribution shift' or `machine drift') whenever ion source species are switched or the ion source is re-started (after overnight turn-off). This means that the past data does not represent the current machine status well. Therefore, we need a strategy that is capable of adopting the distribution shift in a sample-efficient way. Reinforcement learning can be a possible candidate for its ability to adapt to the distribution shift if the shift is not too large. However, its sample efficiency of online adaption to the distribution shift is questionable~\citep{offlineRL}. Uncertainty quantification (UQ) on historical data models can also be a possible solution. However, UQ methods are computationally expensive and highly dependent on the method and hyper-parameters~\citep{randomizedprior}. Instead, we seek to train a deterministic neural network on historical data and use it as a prior mean for the Gaussian Process (GP) surrogate model of Bayesian optimization. Although the scalability problem of GP still remains, the prior mean may help BO to find a good solution quickly in a sample efficient way before the scalability problem becomes an issue. Also, finding a good enough solution in a few steps well aligns with our goal of fast tuning. 

\section{Related Work}
The multi-fidelity Bayesian optimization methods incorporate the historical (or simulational data) into the low-fidelity prior model while building the high-fidelity posterior through online optimization. This method is used for wake-field accelerator tuning~\citep{multiBO}.  However, it suffers from high computational costs due to the scalability issue.

Incorporating the probabilistic prior model into BO while avoiding the scalability problem is possible through augmenting the acquisition function with the prior model~\citep{augmentingAcquisition}. However, in the presence of a significant distribution shift within the historical training data, the reliability of the UQ of the probabilistic prior model is questionable, and the computational overhead still may not be ignorable.

Benchmarking prior mean assisted Bayesian optimization (pmBO) on Rosenbrock and Rastigirin functions while considering distribution shifts within the virtual historical data for training prior mean are presented in our previous work~\citep{pmBO}. A similar idea of using a Neural Network (NN) prior surrogate mean model for efficient BO is also developed in a parallel work~\citep{slac}.

\section{FRIB Front-End historical data model}

\subsection{FRIB Front-End}
FRIB FE consists of a 14 GHz electron cyclotron resonance (ECR) ion source (called ARTEMIS) on a high voltage platform, Low Energy Beam Transport (LEBT), Radio Frequency Resonator (RFQ), and Medium Energy Beam Transport (MEBT). LEBT transport 12 keV/u beam and consists of the chopper, electrostatic quadrupoles, solenoids, electric correctors and etc. MEBT transport 500 keV/u beam and consists of two RF bunchers, quadrupole magnets, 3 beam position monitors (BPM) and etc. 

\subsection{Historical data}
Settings and readings data of various elements are archived for several years. However, as the FRIB started user operation only recently (early 2022), there have been elements location changes, and calibration changes. In addition, the frequent change of ion source species makes the archived data not consistent ( i.e., distribution shifts are persistent within training data as well as possible data from the future machine state). Moreover, hundreds of data dimensions make FE modeling from historical data even more difficult. Nevertheless, we hoped that the prior mean model from the archived data would benefit the BO more than nothing based on our previous benchmark study on toymodels~\citep{pmBO}.

The ion species and charge state that were available to us for the beam test was 36Ar10+. We found that there were 11 days of 36Ar10+ beam operation in the past 2 years. After cleaning out duplicated data, we could obtain 539 data points which are very tiny amounts for hundreds of data dimensions. Nevertheless, there may be a large correlation between features and, importantly, most of the settings are kept the same for the same beam species. Therefore, the historical data may still benefit more than nothing.  

\subsection{Feature selection}
Out of hundreds of features (of data), we selected about 38 input features and 12 output features based on domain expert knowledge and weight activation under Lasso regulerization~\citep{lasso} on the input nodes of NN. The selected input features include solenoids, electric quadrupoles, electric correctors, slit locations, and source parameters. The output features are 3 MEBT BPM's $x, y, \phi$ positions, and signal magnitudes. With the selected features, we trained a NN for the historical data model.

\subsection{Historical data model test}
\begin{figure}
	\centering
	\includegraphics{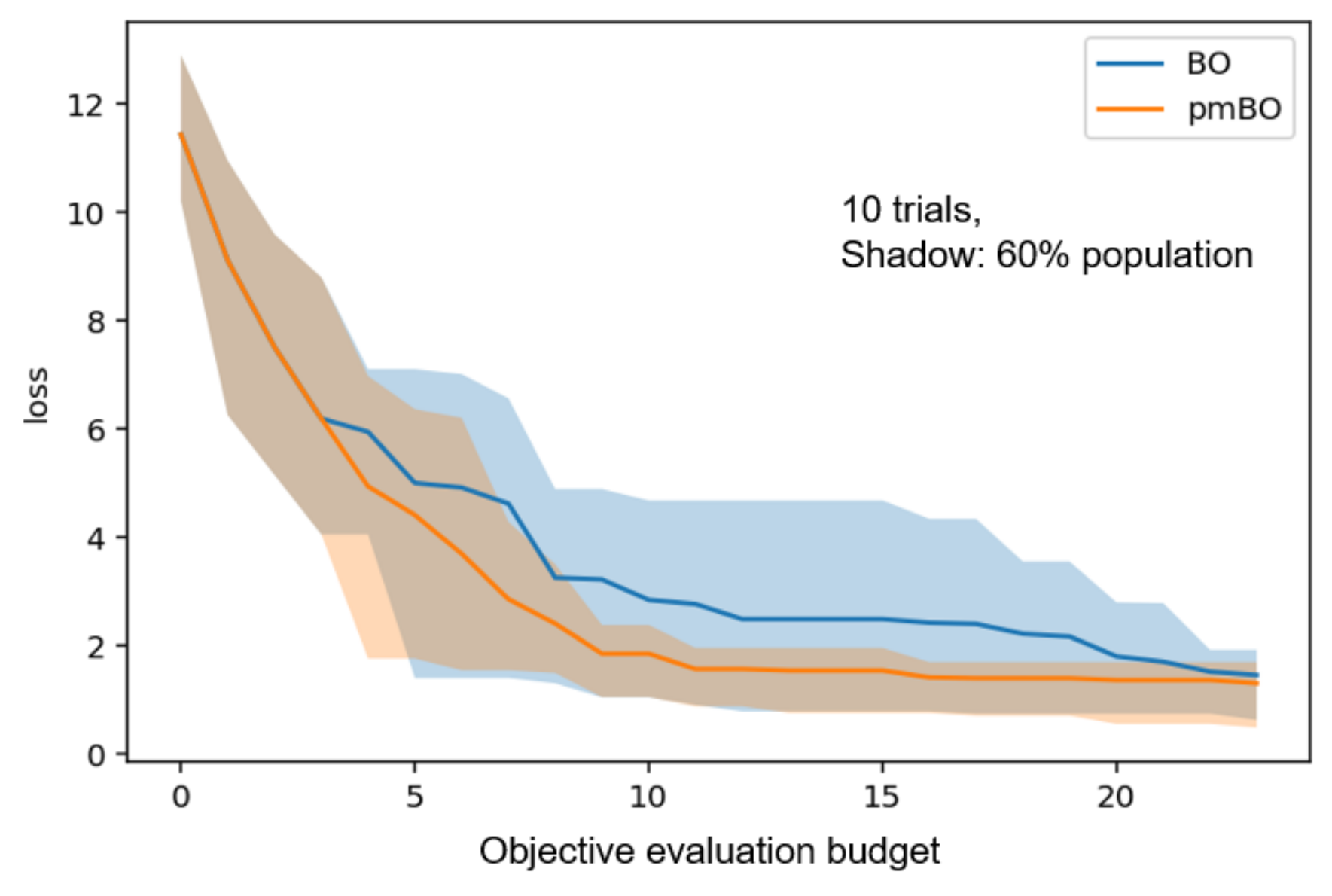}
	\caption{Virtual FE tuning benchmark of pmBO against BO. A small loss means a better objective. And best (out of all the evaluations up to the most recent) objectives are plotted. The `objective evaluation budget' equals the number of pmBO iterations as we queried a single candidate in each BO iteration for this test. The shadowed region includes 80\% upper and 20\% lower loss out of the 10 trials.}
	\label{fig:virtualtest}
\end{figure}

\begin{figure}
	\centering
	\includegraphics{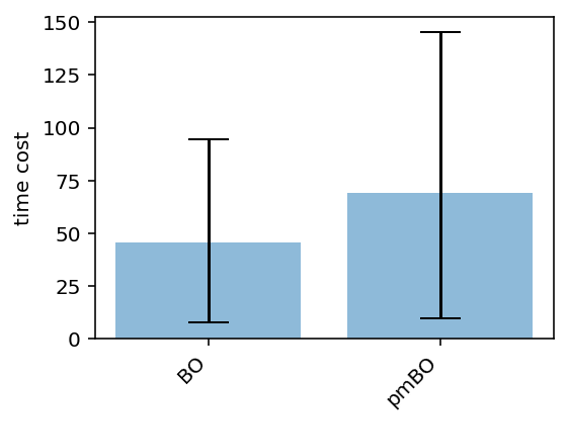}
	\caption{Measured computational complexity comparison in terms of wall time. The error shows 80\% upper and 20\% lower time cost out of the 10 trials.}
	\label{fig:virtualtest_timecost}
\end{figure}

In order to simulate the upcoming FE tuning task (that we did and presented in later section~\ref{sec:beamtest}) with pmBO, we trained two historical models using data before 2022-01-24 and data from 2022-01-25. Then use the first historical model as the prior mean model for pmBO and use the latter one to simulate a virtual machine for the test bed. In this test, we choose 5 decision parameters: 2 horizontal, 2 vertical correctors, and 1 solenoid. And constructed an objective function based on $x, y, \phi$ positions and signal magnitudes of virtual BPMs that is the output of the test model. Figure~\ref{fig:virtualtest} shows the benchmark result of pmBO against (vanilla) BO with 10 trials. In each trial, different environmental variables (that are all the input parameters of the NN model other than the decision parameters) are randomly sampled and fixed during the optimization. As can be seen, pmBO outperformed (vanilla) BO with a non-negligible statistical margin. This result gave us confidence that, although the number of dataset seems not enough, the historical model will provide some meaningful prior information better than nothing. Figure~\ref{fig:virtualtest_timecost} shows the statistics of the total (all iteration) time cost of each trial. The computational complexity of pmBO is a little higher than (vanilla) BO due to the expected overhead from the NN prior mean model.

\section{FE tuning with pmBO}\label{sec:beamtest}
Our goal of the FE tuning is the beam centering and maximization of beam transmission as fast as possible. Any regret from losing the chance of finding a better global minimum by stopping the tuning earlier is less concern. For fast tuning, we choose 3 decision parameters: 1 horizontal, 1 vertical corrector, and 1 solenoid. And fix all other input parameters of pre-trained prior mean NN model at the current values of the machine. We constructed an objective function based on  $x, y, \phi$ positions and signal magnitudes of BPMs. 

\subsection{Objective evaluation time cost and batch candiates}

Although the data collection frequency of the MEBT BPM is faster than 100Hz, because the power ramping of decision parameters takes a few seconds, the time cost of the objective evaluation is a few seconds. In order to reduce the BO computational time cost compared to the objective evaluation, we queried 5 candidates in a batch for objective evaluation for the first few BO iteration and reduced the batch size as iteration continues. We used Botorch's~\citep{botorch} capability of querying multiple candidates in a shot while assuring each candidate is not close to each other in the input space~(that is decision parameter space).

\subsection{FE tuning result}
It would be best if we can benchmark pmBO against BO with enough statistics. However, as we had to do the fast tuning without taking too much beam time, we present FE tuning with pmBO only. However, through the investigation of the GP surrogate model evolution along the tuning process, we could observe that the prior mean model provided meaningful information for fast tuning. We are planning to do a better benchmark in the near future if enough beam time is allowed. 

\begin{figure}
	\centering
	\includegraphics[scale=0.8]{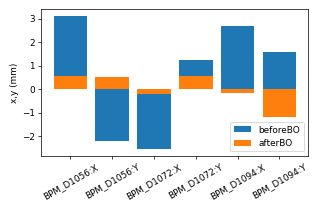}
	\caption{x,y beam position readings before and after FE tuning}
	\label{fig:beforeafterBPMxy}
\end{figure}

\begin{figure}
	\centering
	\includegraphics{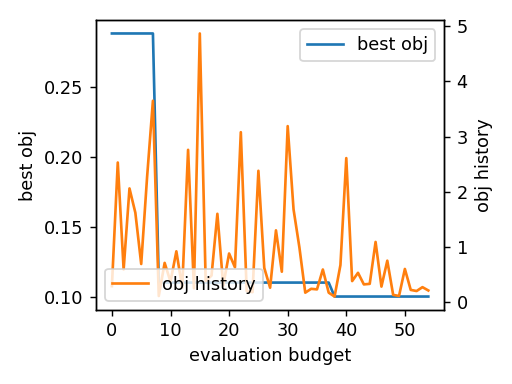}
	\caption{best of the evaluated objective and the objective history}
	\label{fig:obj_history}
\end{figure}

\begin{figure}
	\centering
	\includegraphics[width=\textwidth,keepaspectratio]{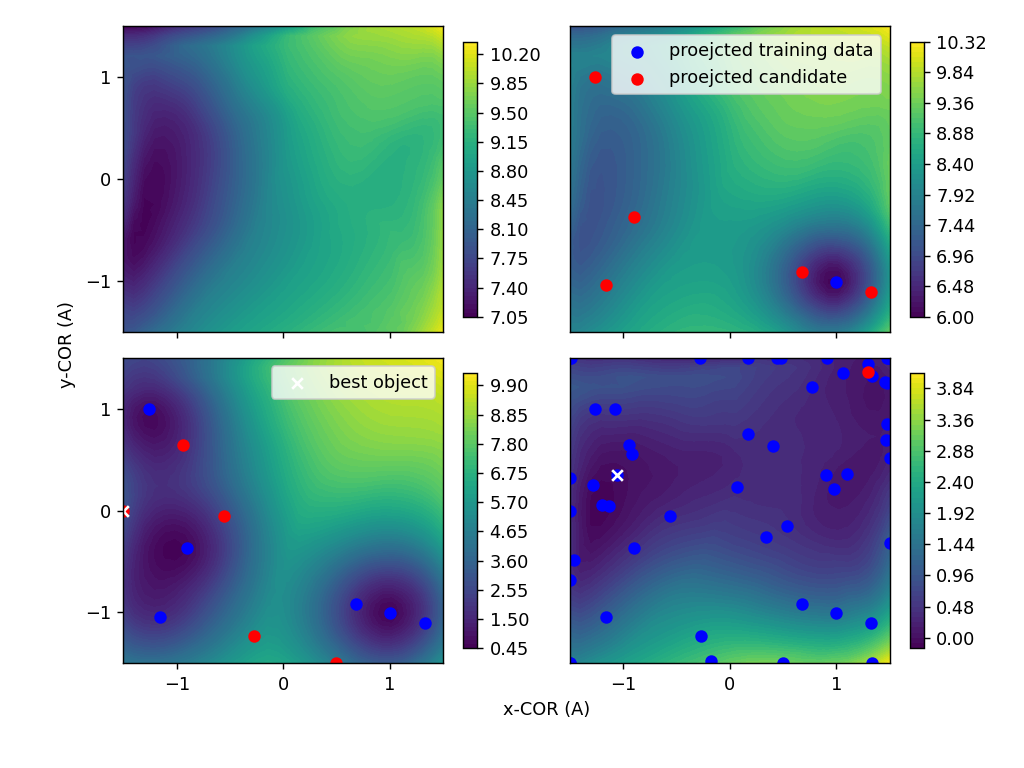}
	\caption{projected prior mean (left top) and GP surrogate model at iteration 1~(right-top), 2~(left-bottom), and the last iteration (right-bottom) onto horizontal and vertical corrector parameter space. The projection is the minimum of model output along the other dimension orthogonal to the projection dimension. The blue points are training data, the red points are queried candidate solutions by pmBO and the white mark is the best objective among them.}
	\label{fig:proejction_36Ar10_0_1_2_last_colorbar}
\end{figure}

Before the start of the FE tuning, the beam transmission was about 23\%, and the transverse beam centroid was mostly larger than $\pm2~(mm)$ from the BPM readings. And after tuning, the 80\% beam transmission could be achieved and the transverse beam centroid was mostly smaller than $\pm1~(mm)$ as shown in Fig~\ref{fig:beforeafterBPMxy}. 
Figure~\ref{fig:obj_history} shows the best of the evaluated objective and the objective history over the course of FE tuning. Note that we found the near convergent solution in the 2nd iteration (recall that we queried 5 candidates in the first few iterations). 
Figure~\ref{fig:proejction_36Ar10_0_1_2_last_colorbar} shows the prior mean (left-top) and evolution of the mean of the GP surrogate model. In order to visualize 3-dimensional input space, we projected the minimum of the predicted objective over the solenoid dimension onto the horizontal and vertical corrector domain. Note that the prior mean model that is trained over the historical data already suggests the region (the dark region on the left side of the plot domain) of the future best objective. This helped the pmBO to provide the candidates the near convergent solution within the 2nd iteration (shown in the left-bottom plot). Note (from the color bar range) also the large offset between the prior mean model and GP prediction at the last iteration. Such offset signifies the distribution shift from the historical data and is adaptively corrected as pmBO proceeds. 

\section{Conclusion}
We tested prior mean model-assisted Bayesian Optimization for FRIB FE tuning. The NN prior mean model is trained over the historical data. Although the historical data is subject to persistent distribution shift, and only a handful of datasets could be retrieved after cleaning out duplicates, outliers, etc, the prior mean provided meaning information for FE tuning (as can be seen from Fig~\ref{fig:proejction_36Ar10_0_1_2_last_colorbar}). This allowed us to reach a near convergent solution in the 2nd iteration (5 candidate solutions in each iteration).

\section*{\centering ACKNOWLEDGEMENTS}
This work supported by the Director of the Office of Science of the US Department of Energy under Cooperative Agreement DE-SC0000661, the State of Michigan and Michigan State University

\iffalse
	\begin{table}
		\caption{Sample table title}
		\centering
		\begin{tabular}{lll}
			\toprule
			\multicolumn{2}{c}{Part}                   \\
			\cmidrule(r){1-2}
			Name     & Description     & Size ($\mu$m) \\
			\midrule
			Dendrite & Input terminal  & $\sim$100     \\
			Axon     & Output terminal & $\sim$10      \\
			Soma     & Cell body       & up to $10^6$  \\
			\bottomrule
		\end{tabular}
		\label{tab:table}
	\end{table}
	
	\begin{itemize}
		\item Lorem ipsum dolor sit amet
		\item consectetur adipiscing elit.
		\item Aliquam dignissim blandit est, in dictum tortor gravida eget. In ac rutrum magna.
	\end{itemize}
\fi

\bibliographystyle{unsrtnat}
\bibliography{template}% Produces the bibliography via BibTeX.

\begin{thebibliography}{8}
\providecommand{\natexlab}[1]{#1}
\providecommand{\url}[1]{\texttt{#1}}
\expandafter\ifx\csname urlstyle\endcsname\relax
  \providecommand{\doi}[1]{doi: #1}\else
  \providecommand{\doi}{doi: \begingroup \urlstyle{rm}\Url}\fi

\bibitem[Levine et~al.(2020)Levine, Kumar, Tucker, and Fu]{offlineRL}
Sergey Levine, Aviral Kumar, George Tucker, and Justin Fu.
\newblock Offline reinforcement learning: Tutorial, review, and perspectives on
  open problems, 2020.
\newblock URL \url{https://arxiv.org/abs/2005.01643}.

\bibitem[Osband et~al.(2018)Osband, Aslanides, and Cassirer]{randomizedprior}
Ian Osband, John Aslanides, and Albin Cassirer.
\newblock Randomized prior functions for deep reinforcement learning, 2018.
\newblock URL \url{https://arxiv.org/abs/1806.03335}.

\bibitem[Pousa et~al.(2022)Pousa, Hudson, Huebl, Jalas, Kirchen, Larson, Lehé,
  de~la Ossa, Thévenet, and Vay]{multiBO}
Á.~Ferran Pousa, S.T.P. Hudson, A.~Huebl, S.~Jalas, M.~Kirchen, J.M. Larson,
  R.~Lehé, A.~Martinez de~la Ossa, M.~Thévenet, and J.-L. Vay.
\newblock {Multitask Optimization of Laser-Plasma Accelerators Using Simulation
  Codes with Different Fidelities}.
\newblock In \emph{Proc. IPAC'22}, number~13 in International Particle
  Accelerator Conference, pages 1761--1764. JACoW Publishing, Geneva,
  Switzerland, 07 2022.
\newblock ISBN 978-3-95450-227-1.
\newblock \doi{10.18429/JACoW-IPAC2022-WEPOST030}.
\newblock URL \url{https://jacow.org/ipac2022/papers/wepost030.pdf}.

\bibitem[Hvarfner et~al.(2022)Hvarfner, Stoll, Souza, Lindauer, Hutter, and
  Nardi]{augmentingAcquisition}
Carl Hvarfner, Danny Stoll, Artur Souza, Marius Lindauer, Frank Hutter, and
  Luigi Nardi.
\newblock Augmenting acquisition functions with user beliefs for bayesian
  optimization, 2022.
\newblock URL \url{https://arxiv.org/abs/2204.11051}.

\bibitem[Hwang et~al.(2022)Hwang, Maruta, Plastun, Fukushima, Zhang, Zhao, and
  Ostroumov]{pmBO}
Kilean Hwang, Tomofumi Maruta, Alexander Plastun, Kei Fukushima, Tong Zhang,
  Qiang Zhao, and Peter Ostroumov.
\newblock {Beam Tuning at the FRIB Front End Using Machine Learning}.
\newblock In \emph{Proc. IPAC'22}, number~13 in International Particle
  Accelerator Conference, pages 983--986. JACoW Publishing, Geneva,
  Switzerland, 07 2022.
\newblock ISBN 978-3-95450-227-1.
\newblock \doi{10.18429/JACoW-IPAC2022-TUPOST053}.
\newblock URL \url{https://jacow.org/ipac2022/papers/tupost053.pdf}.

\bibitem[Xu et~al.(2022)Xu, Edelen, and Roussel]{slac}
Connie Xu, Auralee Edelen, and Ryan Roussel.
\newblock Neural network surrogate priors for efficient bayesian optimization,
  2022.
\newblock URL \url{https://indico.bnl.gov/event/16158/contributions/69991/}.

\bibitem[Muthukrishnan and Rohini(2016)]{lasso}
R~Muthukrishnan and R~Rohini.
\newblock Lasso: A feature selection technique in predictive modeling for
  machine learning.
\newblock In \emph{2016 IEEE International Conference on Advances in Computer
  Applications (ICACA)}, pages 18--20, 2016.
\newblock \doi{10.1109/ICACA.2016.7887916}.

\bibitem[Balandat et~al.(2019)Balandat, Karrer, Jiang, Daulton, Letham, Wilson,
  and Bakshy]{botorch}
Maximilian Balandat, Brian Karrer, Daniel~R. Jiang, Samuel Daulton, Benjamin
  Letham, Andrew~Gordon Wilson, and Eytan Bakshy.
\newblock Botorch: A framework for efficient monte-carlo bayesian optimization.
\newblock 2019.
\newblock \doi{10.48550/ARXIV.1910.06403}.
\newblock URL \url{https://arxiv.org/abs/1910.06403}.

\end{thebibliography}





% \end{thebibliography}

\end{document}